\documentclass[journal=nalefd,manuscript=article]{achemso}
\setkeys{acs}{usetitle = false}
\usepackage[version=3]{mhchem} % Formula subscripts using \ce{}
\usepackage[T1]{fontenc}       % Use modern font encodings
\usepackage{graphicx}
\usepackage{amssymb}
\usepackage{amsmath}
\usepackage[ansinew]{inputenc}
\usepackage{bm}
\usepackage{subfigure}
\usepackage{graphpap}
\usepackage{natmove}
\usepackage{multirow}
\usepackage{cleveref}
\usepackage{color}

\definecolor{Blu}{rgb}{0.00,0.00,1.00}
\definecolor{Red}{rgb}{1.00,0.00,0.00}
\definecolor{Green}{rgb}{0.00,0.60,0.00}
\author{Michael C. Chong}

\author{Nasima Afshar-Imani}

\author{Fabrice Scheurer}
\affiliation{Institut de Physique et Chimie des Mat\'eriaux de Strasbourg, UMR 7504 (CNRS -- Universit\'e de Strasbourg), 67034 Strasbourg, France}
\author{Claudia Cardoso}

\author{Andrea Ferretti}

\author{Deborah Prezzi}
\email{deborah.prezzi@nano.cnr.it}
\affiliation{CNR-Nanoscience Institute, S3 Center, 41125 Modena, Italy}

\author{Guillaume Schull}
\email{schull@ipcms.unistra.fr}
\affiliation{Institut de Physique et Chimie des Mat\'eriaux de Strasbourg, UMR 7504 (CNRS -- Universit\'e de Strasbourg), 67034 Strasbourg, France}
\date{\today}
\title{Bright electroluminescence from single graphene nanoribbon junctions}
\begin{document}
%\begin{tocentry}
%The surrounding frame is 9\,cm by 3.5\,cm, which is the maximum
%permitted for  \emph{Journal of the American Chemical Society}
%graphical table of content entries. The box will not resize if the
%content is too big: instead it will overflow the edge of the box.

%This box and the associated title will always be printed on a
%separate page at the end of the document.
%\includegraphics[height=3.5cm]{Figures/TOC.jpg}
%\end{tocentry}

\begin{abstract}
Thanks to their highly tunable band gaps, graphene nanoribbons (GNRs) with atomically precise edges are emerging as mechanically and chemically robust candidates for nanoscale light emitting devices of modulable emission color. While their optical properties have been addressed theoretically in depth, only few experimental studies exist, limited to ensemble measurements and without any attempt to integrate them in an electronic-like circuit.
Here we report on the electroluminescence of individual GNRs suspended between the tip of a scanning tunneling microscope (STM) and a Au(111) substrate, constituting thus a realistic opto-electronic circuit. Emission spectra of such GNR junctions reveal a bright and narrow band emission of red light, whose energy can be tuned with the bias voltage applied to the junction, but always lying below the gap of infinite GNRs. Comparison with {\it ab initio} calculations indicate that the emission involves electronic states localized at the GNR termini. Our results shed light on unpredicted optical transitions in GNRs and provide a promising route for the realization of bright, robust and controllable graphene-based light emitting devices.
\end{abstract}

\newpage

While graphene is a promising material for a number of electronic applications, the absence of an optical gap limits its use for light emitting devices~\cite{wang2014}. 
Lateral confinement to form nanometer-wide graphene nanoribbons (GNRs) allows one to combine many of the outstanding properties of graphene with the presence of a sizeable gap~\cite{castroneto2009}. The reach of atomic precision over the structure~\cite{Cai2010,cai2014,Narita2014,Ruffieux2016} further opens the way to the production of materials with widely tunable properties, where electronic and optical gaps are predicted to depend on both quantum confinement and edge morphology~\cite{Son2006,Yang2007,Yang2008,Prezzi2008,Prezzi2011,Osella2012,Yazyev2013}, and can be further tuned by edge functionalization~\cite{cai2014,Cocchi2011}. 

Experimentally, the electronic gap of GNRs has been explored for a variety of structures\cite{Ruffieux2012,Linden2012,chen2013,cai2014,kimouche2015,zhang2015,Ruffieux2016,talirz2017} and their optical characterization has been addressed in several cases \cite{Jensen2013,Narita2014,Narita2014b,Denk2014,Lim2015,Soavi2016,Senkovskiy2017,Zhao2017,Candini2017}. The latter involves ultraviolet-visible-near infrared absorbance spectroscopy~\cite{Narita2014,Narita2014b,Lim2015,Soavi2016,Zhao2017}, resonance Raman spectroscopy~\cite{Senkovskiy2017}, reflectance difference spectroscopy~\cite{Denk2014}, as well as time-resolved spectroscopies~\cite{Soavi2016,Jensen2013}. Nonetheless, the investigation of their emission properties, which is relevant in the scope of future electroluminescent nanoscale devices, has been limited so far to a few ensemble measurements at room temperature, only revealing weak and featureless emission spectra~\cite{Zhao2017,Senkovskiy2017}. A better understanding of the intrinsic luminescence properties of GNRs, which may include the effect of specific atomic-scale defects, requires instead to go beyond ensemble measurements and to address the emission at the level of individual ribbons -- a field that is totally unexplored to date. 

\begin{figure}
\centerline{\includegraphics[width=\textwidth]{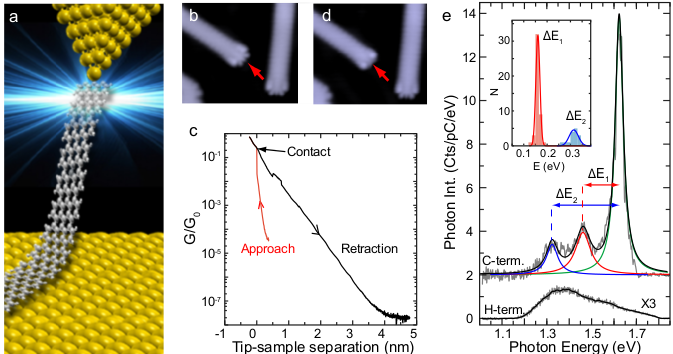}}
\caption{(a) Schematics of the experimental configuration. (b) STM image of H-terminated 7-AGNRs on Au(111) ($8.5\times 7.4$ nm$^2$, $V$ = 0.05 V, $I$ = 0.1 nA). (c) Normalized conductance $G/G_{0}$ \textit{vs} tip-sample distance $z$ for a 7-AGNRs suspended in the junction (black line). The red curve corresponds to the approach of the metallic STM tip to the extremity of the 7-AGNR, in the position marked by an arrow in (b) (V = 0.1 V). (d) STM image of the same area than in (b) after dehydrogenation of the central carbon atom of the ribbon terminus marked by an arrow. (e) STM-induced light emission (STM-LE) spectra of the suspended ribbon when H-terminated (bottom curve, magnified by a factor 3, $z$ = 3.2 nm , $V$ = 1.8 V, $I$ = 14.8 nA, acquisition time $t$ = 60 s) and when C-terminated (top curve, vertically shifted, $z$ = 3.2 nm, $V$ = 1.7 V, $I$ = 0.4 nA, $t$ = 60 s). The inset shows the distribution of the energy shifts of the two low-intensity features from the main peak, {\it i.e.}, $\Delta E_1$ and $\Delta E_2$, obtained from measurement with different C-terminated  junctions.
}\label{fig1}
\end{figure}

Here, we apply a novel approach \cite{Reecht2014,Chong2016,Chong2016a} that consists in lifting an individual GNR between the tip and the substrate of a STM \cite{Koch2012} to study its electroluminescent properties. In this geometry, one extremity of the GNR is lying flat on a Au(111) surface and the other is connected to a gold-capped tip (\Cref{fig1}a). This strongly reduces the coupling between the GNR and the electrodes that otherwise quenches the emission. Here, we use 7-atom-wide armchair GNRs (7-AGNRs) as they present an optical gap in the visible spectrum ($\approx$ 2 eV),\cite{Denk2014} and can be easily obtained by on-surface synthesis \cite{Cai2010,Koch2012,Lit2013,Talirz2013,Denk2014}.
The electroluminescence spectrum of such a junction reveals to be strongly dependent on the nature of the GNR-tip connection. In the case of a covalent bond, an intense and sharp (FWHM = 40 meV) luminescence peak arises, whose energy varies with voltage and tip-sample distance. Extrapolating to an unbiased junction, a transition energy of $\approx$ 1.16 eV is obtained,  significantly lower than the optical gap ($\approx$ 2 eV) reported for 7-AGNRs on Au \cite{Denk2014}. We performed {\it ab initio} calculations at the $GW$ plus Bethe-Salpeter equation ($GW$-BSE) level to simulate the optical spectra of finite-size GNRs. On this basis, the emission band can be assigned to a transition between a state localized at a GNR extremity and a state delocalized along the ribbon.

%\section{Results and discussion}
%%%%%%%%%% Experiment %%%%%%%%%%
%

%EXP DETAILS. The ribbons were synthesized on-surface by annealing (at 670 K) a clean and flat Au(111) single-crystal covered with 0.5 ML of 10,10'-dibromo-9,9'-bianthryl deposited by sublimation under vacuum. The sample was then transferred in situ to the analysis chamber and introduced in an Omicron STM operating at 4 K at a base pressure of 5.10 11 mbar. Chemically etched tungsten tips were treated under vacuum to remove the top oxide layers and, as a final step of their preparation, gently indented in the Au(111) substrate to cover them with gold. The photons emitted at the ribbon junction were collected using a setup that has been described previously [12].

%\subsection{STM-induced light emission from GNR junctions}
\Cref{fig1}b shows a STM image of a typical 7-AGNR polymerized on a Au(111) surface \cite{Cai2010}. The first step of our experiment consists in lifting such a GNR in the STM junction. To do so, the tip of the STM is located on top of the GNR extremity and approached up to the formation of a contact. The tip is then retracted from the surface with the GNR attached to its apex. The junction conductance is monitored during the whole procedure (see \Cref{fig1}c). An exponential decay of the conductance with the tip retraction distance ($i.e.,$ tip-sample separation, $z$) is observed (black curve), with a decay rate much smaller than the increase rate observed during the approach procedure (red curve). This is due to the larger conductivity of the GNR compared to vacuum, and attests for the success of the lifting procedure~\cite{Koch2012}. 
The bottom curve in \Cref{fig1}e is a typical electroluminescence spectrum of a 7-AGNR lifted from the Au(111) surface ($z = 3.2$ nm, $V = 1.8$ V): It reveals a featureless and weak emission spectrum that resembles a typical plasmon spectrum recorded with a pure metal-metal junction~\cite{Berndt1991}.

A very different behavior is observed for 7-AGNRs where the extremity lifted by the tip has been dehydrogenated beforehand. This is done, prior to the lifting experiment, by locating the tip on top of the GNR termination and ramping the voltage to 4 V at a constant current of 10 nA (see Supporting Information for details). The STM image taken after this procedure (\Cref{fig1}d) shows the characteristic pattern of a dehydrogenated terminus (hereafter called C-terminated, to be compared with the H-terminated case discussed previously), which consists of two bright spots separated by a depression~\cite{Talirz2013,Lit2013}. The corresponding optical spectrum in the top panel of \Cref{fig1}e (recorded with the same conditions as for the H-terminated GNR) reveals this time a rather sharp (FWHM = 40 meV) and intense emission peak at 1.6 eV, as well as two faint, equally-spaced lower-energy peaks, located at 161 ($\pm 9$) meV and 314 ($\pm 20$) meV from the main peak, respectively. Note that the peak shifted by 161 meV is systematically observed and its energy presents an extremely low variability, whereas the second peak is weaker and its energy slightly varies from measurement to measurement (see the inset of \Cref{fig1}e). We believe that this effect is due to variations of the tip-apex structure which slightly affect the electronic and plasmonic properties of the junctions. The two shifted peaks are assigned to the first and second harmonics of an optical vibronic mode. While several modes are observed in this energy region \cite{Cai2010}, we tentatively assign the observed replicas to 
the most intense vibronic mode (D mode) observed at about 1340 cm$^{-1}$ (166 meV) in resonance Raman spectra of 7-AGNRs on Au(111)~\cite{Shiotari2014}. 
The sharpness of the emission and the presence of vibronic features are incompatible with a purely plasmonic \cite{Sch08} or a thermal emission \cite{Duck2015}, and strongly suggest a radiative transition intrinsic to the GNR. While the overall quantum efficiency of the emission process remains relatively low ($10^{-5}$ photon/electron), the ability of the suspended GNRs to withstand elevated currents (up to 40 nA) allows for an intense emission of light that can amount up to $10^7$ photon/s ($\approx$ 1 pW at 30 nA excitation current). For comparison, this is 100 times larger than the largest emission measured with previous molecular scale optoelectronic devices~\cite{Chong2016}, and a factor of $\approx 5$ more intense than the purely plasmonic emission measured in the absence of a GNR with the same current and voltage (see Supporting Information). Related to the size of the emitting area ($\approx$ 1 nm$^{2}$), the emission of the nanoribbon junction is as intense as the one reported for bright light emitting devices made of carbon nanotubes~\cite{Chen2005}.

%\subsection{Voltage dependence of the light emission spectra}

\begin{figure}
\centerline{\includegraphics[width=\textwidth]{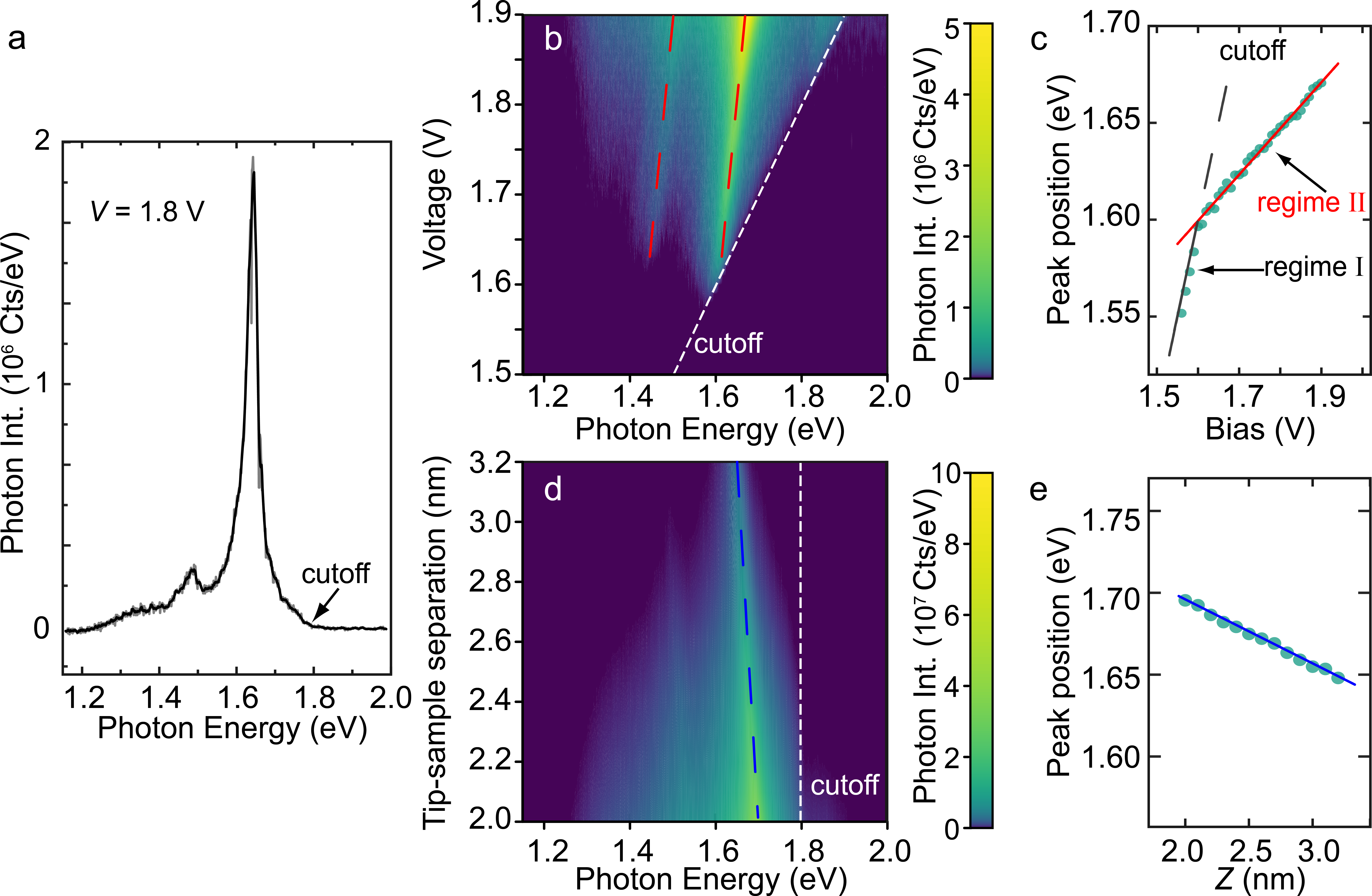}}
\caption{(a) STM-LE spectrum of a suspended C-terminated ribbon ($z$ = 3.2 nm, $V$ = 1.8 V, $I$ = 3.7 nA, $t$ = 18 s). (b, d) Voltage ($z$ = 3.2 nm) and tip-sample distance ($V$ = 1.8 V) dependencies of the STM-LE spectra, respectively. (c, e) Energy shift of the main emission line with $V$ and $z$, respectively.
}\label{fig2}
\end{figure}

Having discussed the main characteristics of the electroluminescence spectrum, we now turn to its voltage dependence (\Cref{fig2}b). For $V \le 1.53$ V, no emission is detected. Above this threshold, one observes the appearance of both the main peak and the vibronic resonances. The energies of the peaks shift to higher values upon increasing voltage, whereas the energy separation between the redshifted resonances and the main emission line remains constant, as expected for excitations of vibronic origin. Applying higher voltages ($V \ge$ 1.9 V) did not reveal other intense spectral features but consistently resulted in junction instabilities and eventually in the rupture of C-C or C-H  bonds within the suspended GNR. An extremely weak and featureless spectrum is measured for the opposite polarity (see Supporting Information), highlighting the unipolar character of the emission.
\Cref{fig2}c displays a plot of the energy shift of the main peak maximum as a function of voltage. In a first regime (1.53 V $\le$ V $\le$ 1.60 V), the shift follows the quantum cut-off line (eV = $h\nu$), as no photon can be emitted above the electron excitation energy. For $V \ge 1.6$ V, a linear shift of 0.23 eV/V of the main emission line is evidenced. A similar behavior is observed in the tip-sample distance dependence of the optical spectrum (see \Cref{fig2}d), where the peak maximum shifts to lower energy with tip retraction (\Cref{fig2}e). 

With respect to the use of nanoribbon junctions as nanoscale optoelectronic devices, these observations provide an interesting opportunity to tune, in-situ, the color of the emission. In addition, they raise intriguing questions regarding the nature of both the optical excitation and the emission mechanism. First of all, one should determine the intrinsic energy of the optical transition, $i.e.,$ in the absence of an applied bias. Assuming a constant linear shift of 0.23 eV/V of the emission maximum, one can deduce an emission energy of $1.16 \pm 0.08$ eV at $V = 0$ V. This is far from the value ($h\nu \approx 2$ eV) reported for the optical absorption onset of 7-AGNRs by reflectance difference spectroscopy~\cite{Denk2014} and resonance Raman measurements~\cite{Senkovskiy2017}, % gap of infinitely long 7-AGNRs~\cite{Denk2014}, where the lowest energy 
which is attributed to the lowest energy excitonic state involving transitions between the last valence and first conduction bands~\cite{Denk2014}. This is also different from the broad luminescence emission reported at about 1.8 eV, and caused by sp$^3$ defects in 7-AGNRs~\cite{Senkovskiy2017}. 

According to calculations for infinite, non-defected GNRs,~\cite{Denk2014} there is no electronic state in the gap that may lead to an emission below the optical onset. Conversely, for finite length ribbons, scanning tunneling spectroscopic (STS) measurements revealed the presence of localized states, also called Tamm states, at the termini of 7-AGNRs~\cite{Koch2012,Lit2013,Wang2016a}. For GNRs directly adsorbed on a metallic substrate, these states show up as a resonance near the Fermi level in the $dI/dV$ curves, and are responsible for the characteristic ``finger-like'' patterns observed in the STM images of GNRs with H- and C-terminated extremities (see \Cref{fig1}b-c, Supporting Information and Ref.~\cite{lijas2013}). In the case of GNRs partially decoupled by an insulating salt layer, the Tamm states appear spin-split with an energy gap of 1.9 eV almost independent of the GNR length~\cite{Wang2016a}. 

\begin{figure}
\centerline{\includegraphics[width=0.45\textwidth]{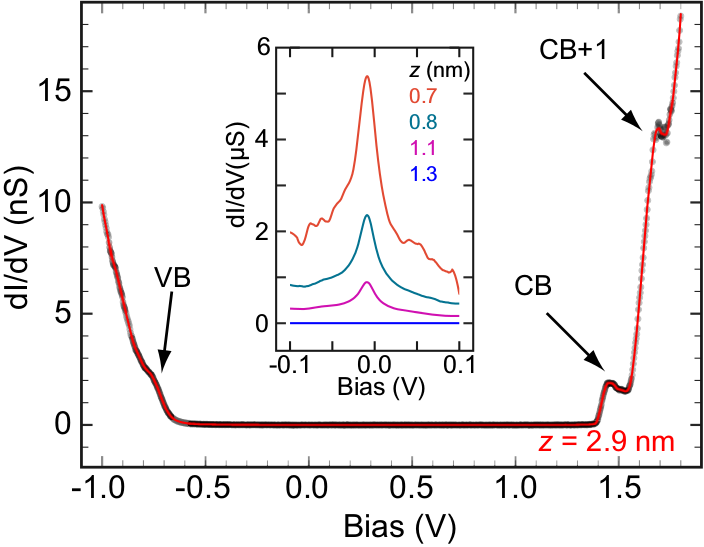}}
\caption{Conductance dI/dV spectra of a suspended 7-AGNR. The tip-sample separation is set to $z = 2.9$ nm. The inset shows a resonance near the Fermi level ($z = 0.7$ to 1.3 nm) corresponding to the Tamm state located at the ribbon extremity connected to the tip. The Tamm state decays for larger tip-sample separations. For large $z$ values ({\it e.g.}, $z = 1.3$ nm) the contribution of the Tamm state is too weak to be measured and only the delocalized valence and conduction bands of the GNR are observed.
}\label{fig3}
\end{figure}

We thus inspected the $dI/dV$ spectrum acquired on a suspended ribbon looking for the presence of midgap states. \Cref{fig3} clearly reveals resonances at $V = -0.8$ V and $V = 1.5$ V. Assuming that the voltage drops mostly at the tip-GNR junction \cite{Koch2012}, these states can be assigned to the valence band (VB) top and conduction band (CB) bottom of the 7-AGNR, respectively. This assumption is supported by the good agreement with the VB top and CB bottom energies ($\approx$ -0.8 eV and $\approx$ 1.5 eV, respectively) as measured by STS for GNRs lying flat on the Au surface~\cite{Sode2015}. In addition, during the first steps of the lifting procedure of a C-terminated 7-AGNR, we often observe a resonance near the Fermi level (see inset of \Cref{fig3}), that is assigned to the Tamm state. Its signature progressively disappears as the tip is further retracted from the surface. This may be expected considering the spatial localization of the state at the GNR terminus, which limits the contribution of the state to the junction conductance at large tip-sample distance. 
Our conductance measurements therefore confirm the presence of an end state (Tamm state) in the gap also for the lifted configuration, in addition to the presence of states delocalized along the GNR junction (VB and CB). Moreover, at variance with the case of GNRs on salt, only one feature is observed, located near the Fermi level. Therefore, despite the realization of a partial decoupling of the GNR from the metallic substrate in the lifted configuration, as demonstrated by the sharp light emission, the coupling with the tip seems to affect the end state significantly.

%%%%%%%%%% Theory %%%%%%%%%%
%
%Because previous optics studies mainly focused on infinitely long ribbons, the possible impact of the Tamm state on optical transitions has been so far overlooked. %

%\subsection{Optical response of finite 7-AGNRs from first principles}

%In order to assess the possible impact of the Tamm state on the optical activity of finite 7-GNRs, we resort to ab-initio many-body perturbation theory calculations in the $GW$-BSE framework (see Methods). This theoretical approach, which has been successfully applied to the case of infinite GNRs~\cite{Denk2014,Prezzi2008}, takes electron-electron interactions into account and is designed for an accurate description of both fundamental gap and optical response. Before that, we need however to understand the influence of the STM tip on the electronic properties of the GNR. We thus simulate from first principles the junction of a finite C-terminated 7-AGNR in contact with the apex of a gold cluster (see \Cref{fig4}a) at the density-functional-theory (DFT) level. In order to single out only the effect of the terminus in contact with gold and avoid any spurious end-to-end interactions, one of the two ends is terminated with two additional C-rings (see \Cref{fig4}a), thus removing the corresponding Tamm states {\col CC: suppressed in the suspended GNR due to ...}. Further details are reported in Supporting Information.

\begin{figure}
%\centerline{\includegraphics[width=0.45\textwidth,clip]{Figures/figure4v5.pdf}}
\centerline{\includegraphics[width=0.9\textwidth,clip]{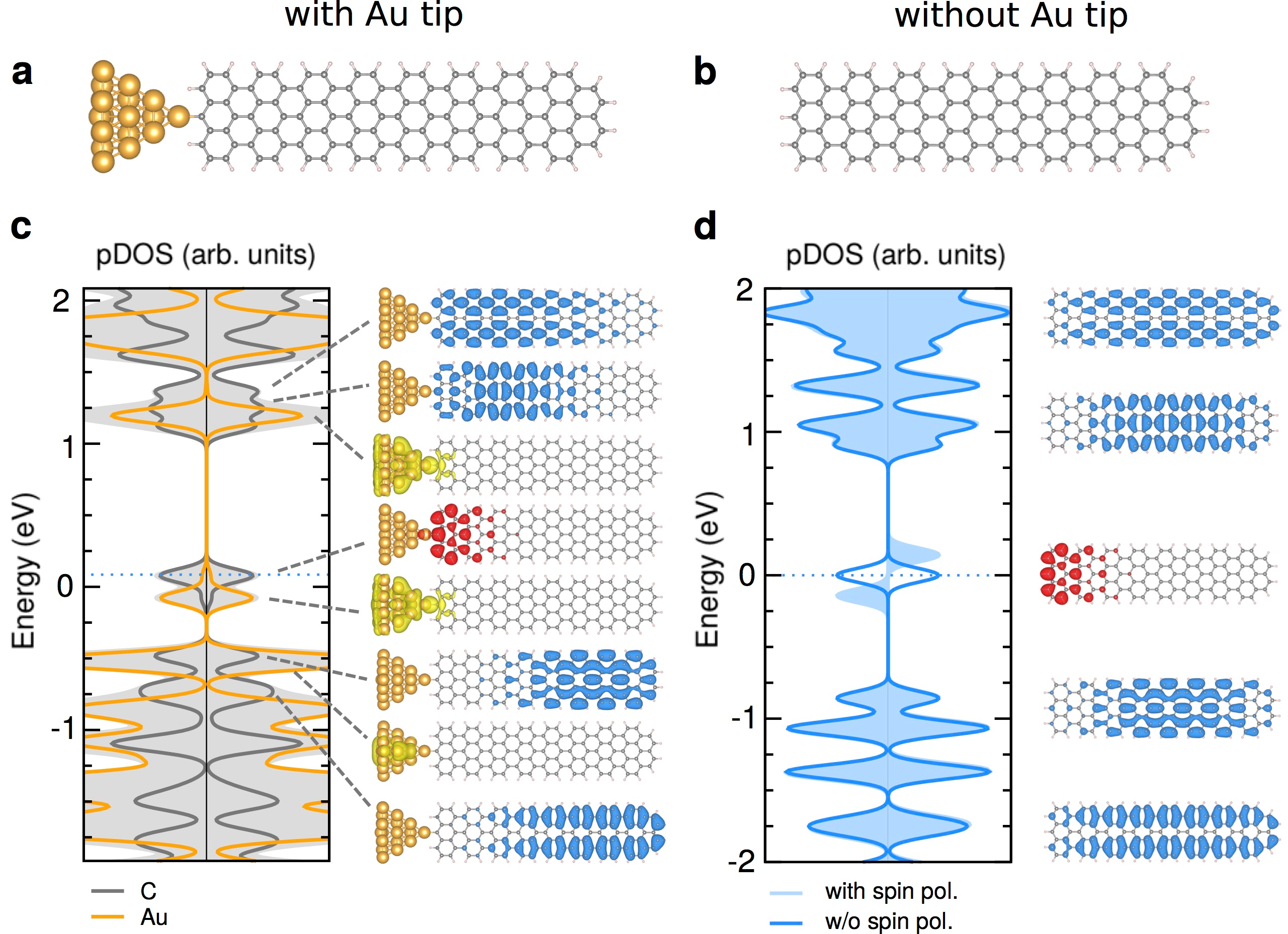}}
%\caption{(a) Ball-and-stick model of a (7,16)-AGNR connected to a 20-atom gold tip. (b) PDOS for the self-standing (7,16)-AGNR, showing both spin-polarized (light blue shaded area) and -unpolarized solutions (blue solid line). (c) PDOS for the GNR-tip junction displayed in panel (a). The total DOS is in grey, and the DOS projected on the C and Au atoms are in dark grey and orange, respectively. }\label{fig4}
\caption{Ball-and-stick model of a (7,16)-AGNR (a) connected to a 20-atom gold tip or (b) terminated with an H atom. (c) DFT-PDOS for the GNR-tip junction displayed in panel (a), together with a few representative molecular orbitals around the Fermi level. The total DOS is in grey, and the DOS projected on the C and Au atoms are in dark grey and orange, respectively. (d) DFT-PDOS for the self-standing (7,16)-AGNR, showing both spin-polarized (light blue shaded area) and -unpolarized solutions (blue solid line). A few representative molecular orbitals around the Fermi level are illustrated also for the isolated system.}\label{fig4}
\end{figure}

In order to better understand the influence of the STM tip on the electronic levels of the GNR, we simulate from first principles the junction of a finite C-terminated 7-AGNR in contact with the apex of a gold cluster (\Cref{fig4}a), and compare its properties with those of its isolated counterpart (\Cref{fig4}b). Note that the free end (i.e. the one not in contact with Au) is terminated with two additional C-rings to remove the corresponding spin-split Tamm states, which are suppressed in real samples by the contact with the gold substrate (see Supporting Information). \Cref{fig4}c,d shows the analysis of the density of states (DOS), as resulting from density-functional-theory (DFT) based calculations. We present here the case of a 3.5-nm-long 7-AGNR [\textit{i.e.} $n=16$ according to the nomenclature proposed in Ref.~\cite{Wang2016a}, and hereafter called (7,16)-AGNR] both in the presence of a 20-atom gold tip (panels a,c) and for the isolated configuration (panels b,d). Convergence with respect to the cluster size and the GNR length is reported in Supplementary Information.
%
%{\DP While starting from a spin-polarized syste the case of isolated 7-AGNRs, with a pair of spin-split Tamm states in the gap (\Cfig4}c, light blue shaded area),} our calculations evidence a complete suppression of the spin polarization {when the GNR is placed in contact with the Au cluster,} as indicated by the projection of the DOS onto the atomic orbitals of C (dark grey line) and Au (orange line) for both spin channels (positive and negative curves).
In the case of the isolated (7,16)-AGNR (\Cref{fig4}b), the zigzag terminus gives rise to a pair of end-localized, spin-split states that lie in the gap defined by the bulk delocalized orbitals (\Cref{fig4}d, light blue shaded area). 
When the GNR is placed in contact with the Au cluster (\Cref{fig4}a), our calculations evidence instead a complete suppression of the spin polarization, as indicated by the projection of the DOS onto the atomic orbitals of C (dark grey line) and Au (orange line) for both spin channels (positive and negative curves) in \Cref{fig4}c.
In particular, the two spin-split Tamm states characterizing the isolated GNR become degenerate and lie near the Fermi level, in agreement with experimental observations. Moreover, the DOS projected on C compares very well to that of the isolated GNR as obtained by spin-restricted DFT calculations, as shown in panel c (dark grey curve) and d (blue solid line) of \Cref{fig4}, 
%. Indeed, the comparison of DOS (Fig.~\ref{fig4}a), molecular orbitals and energy levels (Fig.~\ref{fig4}b) of the Au-GNR junction with those of the isolated GNR indicates only a minor 
indicating only a minor hybridization between the GNR and Au cluster states. This is further corroborated by the one-to-one comparison of the most relevant molecular orbitals, as shown in \Cref{fig4}c,d. These results support the realization of a partial decoupling from Au, while explaining why the Tamm states are not spin-split in our measurements (\Cref{fig3}). 

We next move to the investigation of the optical properties resulting from the (7,16)-AGNR energy level scheme as modified by the presence of the tip. To this end, we need to resort to a higher-level computational approach beyond mean field, able to accurately take into account both electron-electron and electron-hole interactions, which were proven of key importance to correctly describe the nature of optical excitations of nm-wide GNRs~\cite{Prezzi2008,Yang2007} and recover a good agreement with experiments~\cite{Denk2014,Senkovskiy2017,Soavi2016,Denk2017}. Electron-phonon coupling is instead not included in our treatment, being beyond the scope of our study.
The DFT results discussed above allow us to further simplify the system under investigation, suggesting that the optical properties of the GNR in contact with Au can be modelled by starting from the spin-restricted ground state for the isolated GNR, $i.e.,$ neglecting both the spin degree of freedom and the contribution of the gold cluster. This simplification allows us to perform ab-initio many-body perturbation theory calculations in the $GW$-BSE framework (see Methods), which would have been otherwise unfeasible for the complete junction.

\begin{figure}
\centerline{\includegraphics[width=0.9\textwidth]{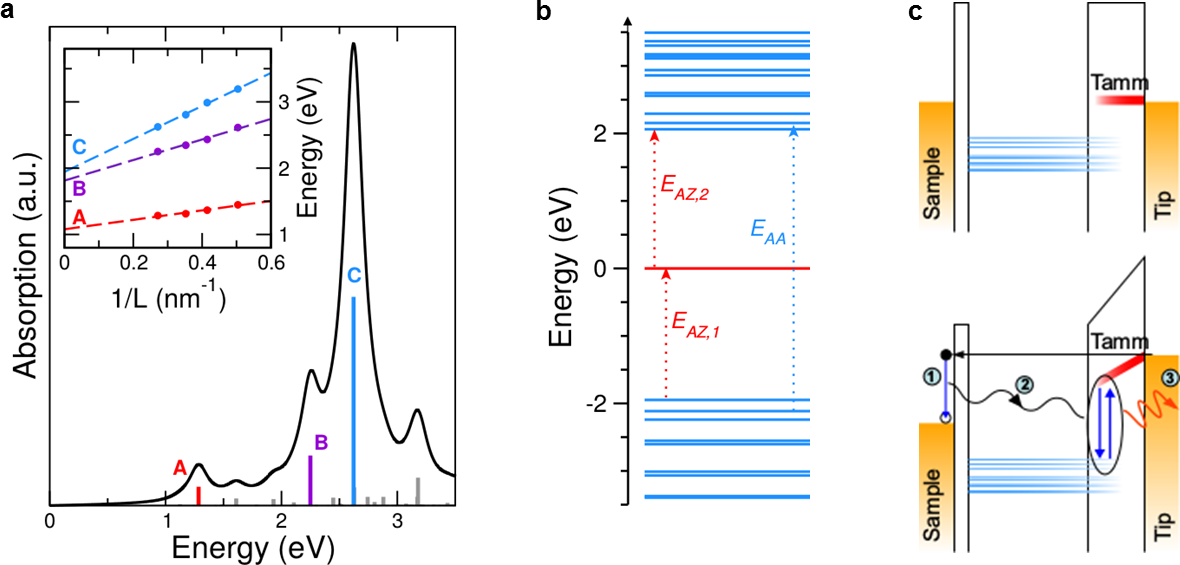}}
\caption{(a) Absorption spectrum computed for the self-standing (7,16)-AGNR according to the $GW$-BSE scheme, starting from the spin-restricted ground state. Inset: Length dependence of the main optical excitations. (b) Energy level scheme for (7,16)-AGNR, as resulting from $GW$ calculations. (c) Sketch of the junction level alignment at $V = 0$ V (top) and $V \approx 1.7$ V (bottom), with illustration of the emission mechanism: an inelastic conduction electron (1) transfers its energy to the GNR (2) that eventually emits a photon (3). For simplicity, only the occupied delocalized states are represented in this sketch, but transitions from the unoccupied delocalized states to the Tamm state may also contribute to the emission.
}\label{fig5}
\end{figure}

\Cref{fig5}a displays the optical absorption spectrum calculated in the $GW$-BSE framework for the (7,16)-AGNR given in \Cref{fig4}b. 
%{\col CC: I would remove the rest of the sentence if we explain the presence of the H when discussing the pDOS above.(}, where the gold cluster is replaced by a H atom to passivate the bond {\col )}. 
The first peak at about 1.3 eV (A) results mainly from transitions between the higher energy occupied bulk states to the singly occupied state localized at the zigzag terminus ($E_{AZ,1}$, see \Cref{fig5}b), and from the latter to the lowest empty bulk state ($E_{AZ,2}$).  
While transitions from/to the Tamm state to/from deeper bulk states characterize also the peaks around 2.2 eV (B), the excitation at about 2.6 eV (C) mainly results from transitions between bulk states ($E_{AA}$, see \Cref{fig5}b).
It is worth noticing that the peak C involving delocalized states is more than half an eV higher in energy than the corresponding excitation for the infinite 7-AGNR \cite{Prezzi2011,Denk2014}, owing to the additional confinement along the ribbon axis. This can be rationalized by looking at the length dependence of the main excitations, as reported in the inset of \Cref{fig5}a for the (7,$n$)-AGNR series, where $n=8-16$. The energy of C is indeed inversely proportional to the GNR length $L$, and varies from 3.2 ($n=8$) to 2.6 eV ($n=16$), while the extrapolated value for $L=\infty$ is 1.9 eV, in excellent agreement with previous calculations~\cite{Prezzi2011,Denk2014} and experimental data from reflectance difference spectroscopy~\cite{Denk2014} and resonance Raman~\cite{Senkovskiy2017}. 
The energy of the A peak shows instead a much less pronounced dependence with length in view of the strongly localized nature of the Tamm state. This leads to an extrapolated value of 1.1 eV for this excitation, in very good agreement with the estimate at zero bias obtained experimentally (1.16 eV). 
We remark that this is the first report of these low-energy excitations, which were not observed in previous absorption experiments. This is probably due to their much lower oscillator strength as compared to that of excitations involving delocalized states resulting from the reduced overlap of the quasi-particle wavefunctions. Their presence can however further explain the unexpectedly low PL recently reported for 7-AGNRs on insulating substrates~\cite{Senkovskiy2017}. 
On the contrary, the higher-energy excitations characterizing the 7-AGNR absorption are not observed in our emission spectra. This is probably due to internal conversions from high- to low-energy excitonic states mediated by vibronic couplings, which are very common in organic systems. These non-radiative decays are generally much faster than radiative processes, preventing the observation of high-energy transitions in emission.

%{\DP I would place the following in the Methods since we have already clarified the issue about the tentative assignment in the exp section:}{\GS As our theoretical approach does not account for vibrational degree of freedom of the GNR, the weak red shifted peaks observed experimentally and associated to vibronic bands (\Cref{fig1}e) are not reproduced in the simulation.}     

%\section{Excitation mechanism}
%%%% Figure 6
After having assigned the main emission feature on the basis of ab initio simulations, it is worth turning to the voltage dependence described in \Cref{fig2}. In fact, the observation of an emission threshold at eV = h$\nu$ (\Cref{fig2}b) suggests an energy transfer between the tunneling electrons and the GNR junction, possibly mediated by the plasmon localized at the tip-sample junction. This excitation mechanism (sketched in \Cref{fig5}c) has been reported on for multilayers of molecules \cite{Dong2010}, molecules separated from metallic surface by thin insulating layers \cite{Imada2017} and, following the same argument, for molecules suspended in a STM junction by decoupling organic wires \cite{Chong2016, Chong2016a}. Here, our calculations suggest that the excitation involves the Tamm state located at the GNR end in contact with the tip and bulk states delocalized along the whole GNR (see energy level scheme in \Cref{fig5}b). Since these states have different spatial localizations, they may experience different energy shifts with voltage (see \Cref{fig2}b). As discussed previously, the voltage drops essentially at the tip-GNR interface, where the Tamm state is localized. Qualitatively, this state remains closer to the tip potential while the delocalized states keep the potential of the Au substrate. This mechanism would induce a net change in the emission gap with voltage, and could explain the energy shifts of the emission with $V$ and $z$ observed in \Cref{fig2}.

%%%%%%%%%%%%%%%%%%%%%%%%%%%%%%%%%%%%%%%%%%%%%%%%%%%%%
% CONCLUSIONS
%%%%%%%%%%%%%%%%%%%%%%%%%%%%%%%%%%%%%%%%%%%%%%%%%%%%%

%\section{Conclusions}

In summary, we report on the first luminescence measurement directly at the level of an individual GNR. The GNR bridges two metallic electrodes forming a real electronic circuit, and its luminescence is excited by electrons, thus providing the most important characteristics for a light emitting device of molecular dimensions. The emission of the GNR junction has a narrow band-width and its colour can be adjusted in-situ by simply tuning the bias voltage. State-of-the-art ab-initio calculations at the $GW$-BSE level reproduce the experimental results with high accuracy and assign the origin of the emission to intra-GNR excitonic transitions involving localized and delocalized electronic states. Thanks to their high robustness and their fair electron to photon conversion efficiency, the emission of the GNR junctions is intense and comparable to the one of highly brilliant light emitting devices made of carbon nanotubes \cite{Chen2005}. Our GNR junctions can therefore be viewed as robust, brilliant and controllable narrow-band light emitting devices, a unique combination that constitute an important step towards the fabrication of realistic optoelectronic components relying on single-molecules as the active element.

%%%%%%%%%%%%%%%%%%%%%%%%%%%%%%%%%%%%%%%%%%%%%%%%%%%%%
% METHOD
%%%%%%%%%%%%%%%%%%%%%%%%%%%%%%%%%%%%%%%%%%%%%%%%%%%%%

\section{Methods}

%%%%%%%%%% Method -- Exp %%%%%%%%%%
{\it Experimental details.}
The ribbons were synthesized on-surface by annealing (at 670 K) a clean and flat Au(111) single-crystal covered with 0.5 ML of 10,10'-dibromo-9,9'-bianthryl deposited by sublimation under vacuum. The sample was then transferred in situ to the analysis chamber and introduced in an Omicron STM operating at 4.5 K at a base pressure of 5.10$^{-11}$ mbar. Chemically etched tungsten tips were treated under vacuum to remove the top oxide layers and, as a final preparation step, gently indented in the Au(111) substrate to cover them with gold. The photons emitted at the ribbon junction were collected using a setup that has been described previously~\cite{Reecht2014}.

%%%%%%%%%% Method -- Theo %%%%%%%%%%
{\it Computational details.}
The ground-state DFT calculations were carried out using the Quantum ESPRESSO package~\cite{gian+09jpcm}. The local density approximation (LDA, Perdew-Zunger parametrization~\cite{perd-zung81prb}) was adopted for the exchange-correlation potential. For the GNR-Au junction, ultrasoft pseudopotentials were used to model the electron-ion interaction. 
The kinetic energy cutoff for the wave functions (charge density) was set to 25 (300) Ry. 
The DFT simulations of the self-standing GNR used for the $GW$-BSE calculations were repeated with norm-conserving pseudopotentials (wave function cutoff set to 80 Ry) as a starting point for the Yambo package~\cite{Marini2009}. GNRs were simulated by employing a supercell size of at least twice the length and width of the studied ribbons, while the distances between the ribbon planes is at least 12~\AA, in order to guarantee no spurious interactions with the system replicas and mimic an isolated system.
The atomic positions within the cell were fully optimized with a force threshold of 0.026 eV/\AA. 

The optical absorption properties were subsequently computed within the framework of many-body perturbation theory~\cite{RevModPhys.74.601}, according to the $GW$-BSE approach. Quasi-particle corrections to the Kohn-Sham eigenvalues were calculated within the $G_0W_0$ approximation for the self-energy operator, where the dynamic dielectric function was obtained within the plasmon-pole approximation. The Coulomb potential was hereafter truncated by using a box-shaped cutoff to remove the long-range interaction between periodic images and simulate isolated systems. The optical absorption spectra were then computed as the imaginary part of the macroscopic dielectric function starting from the solution of the BS equation in order to take into account excitonic effects. The static screening in the direct term was calculated within the random-phase approximation with inclusion of local field effects; the Tamm-Dancoff approximation for the BS Hamiltonian was employed. The aforementioned many-body effects were included using the Yambo code~\cite{Marini2009}. 
%{\DP Note that electron-phonon coupling was not included in our treatment, the latter being beyond the scope of our investigation. Vibronic replicas observed experimentally are thus not expected to be reproduced by the theory.}

%%%%%%%%%%%%%%%%%%%%%%%%%%%%%%%%%%%%%%%%%%%%%%%%%%%%%%%%%%%%%%%%%%
\begin{suppinfo}
{\it Supporting Information} contains: the detailed description of the procedure for the termini dehydrogenation; the analysis of the STM images for H- and C-terminated GNRs, prior and after the lifting experiment, together with a few relevant $dI/dV$ spectra; a discussion of the polarity dependency of the emission spectra; a thorough analysis of the emission intensity; an extended description of the first-principles simulations.   
\end{suppinfo}

%%%%%%%%%%%%%%%%%%%%%%%%%%%%%%%%%%%%%%%%%%%%%%%%%%%%%%%%%%%%%%%%%%
\section{Author Information}

\subsection{ORCID}
Deborah Prezzi: 0000-0002-7294-7450\newline
Andrea Ferretti: 0000-0003-0855-2590\newline
Claudia Cardoso: 0000-0002-9339-8075\newline
Guillaume Schull: 0000-0002-4205-0431 

\subsection{Author Contributions}

G.S. conceived the work. N.A-I., M.C. and G.S. performed the STM experiment. D.P. and A.F. conceived the theoretical study. C.C. carried out the simulations. G.S. and D.P. wrote the manuscript. All authors discussed the results and commented on the manuscript at all stages.

\subsection{Notes}
The authors declare no competing financial interests.

%%%%%%%%%%%%%%%%%%%%%%%%%%%%%%%%%%%%%%%%%%%%%%%%%%%%%%%%%%%%%%%%%%
\begin{acknowledgement}
The authors thank Virginie Speisser, Jean-Georges Faullumel, Michelangelo Roméo and Olivier Cregut for technical support. M.C., N. A-I., F.S. and G.S. acknowledge the Agence National de la Recherche (project SMALL'LED No. ANR-14-CE26-0016-01), the Labex NIE (Contract No. ANR-11-LABX-0058 NIE), the R\'egion Alsace and the International Center for Frontier Research in Chemistry (FRC) for financial support. C.C., A.F. and D.P. acknowledge support from the European Union H2020-EINFRA-2015-1 program (Grant No. 676598, project "MaX - materials at the exascale").
Computational resources were provided by the ISCRA program (Grant No. HP10BO7B2W) and by PRACE via project ``SHINE'' (Grant No. Pra11\_2921) on the Marconi machine at CINECA.
\end{acknowledgement}

%%%%%%%%%%%%%%%%%%%%%%%%%%%%%%%%%%%%%%%%%%%%%%%%%%%%%%%%%%%%%%%%%%
%\bibliography{graphene}
\providecommand{\latin}[1]{#1}
\providecommand*\mcitethebibliography{\thebibliography}
\csname @ifundefined\endcsname{endmcitethebibliography}
  {\let\endmcitethebibliography\endthebibliography}{}

%%%%%%%%%%%%%%%%%%%%%%%%%%%%%%%%%%%%%%%%%%%%%%%%%%%%%
% FIGURES
%%%%%%%%%%%%%%%%%%%%%%%%%%%%%%%%%%%%%%%%%%%%%%%%%%%%%
\newpage

\end{document}